\documentclass[useAMS,usenatbib,usegraphicx]{mn2e}
\usepackage{comment}

\title[Halo Masses and Globular Cluster Systems]{A New Method for Estimating Dark Matter Halo Masses using Globular Cluster Systems}
\author[L. R. Spitler and D. A. Forbes]{Lee R. Spitler$^{1}$\thanks{E-mail: lspitler@astro.swin.edu.au } and Duncan A. Forbes$^{1}$\\
$^{1}$Centre for Astrophysics \& Supercomputing, Swinburne University, Hawthorn, VIC 3122, Australia}

\begin{document}


\pagerange{\pageref{firstpage}--\pageref{lastpage}} \pubyear{2008}

\maketitle

\label{firstpage}

\begin{abstract}
All galaxies are thought to reside within large halos of dark matter, whose properties can only be determined from indirect observations.  The formation and assembly of galaxies is determined from the interplay between these dark matter halos and the baryonic matter they host.  Although statistical relations can be used to approximate how massive a galaxy's halo is, very few individual galaxies have direct measurements of their halo masses.  We present a method to directly estimate the total mass of a galaxy's dark halo using its system of globular clusters.  The link between globular cluster systems and halo masses is independent of a galaxy's type and environment, in contrast to the relationship between galaxy halo and stellar masses.  This trend is expected in models where globular clusters form in early, rare density peaks in the cold dark matter density field and the epoch of reionisation was roughly coeval throughout the Universe.  We illustrate the general utility of this relation by demonstrating that a galaxy's supermassive black hole mass and global X-ray luminosity are directly proportional to their host dark halo masses, as inferred from our new method.
\end{abstract}


\section{Introduction}

Our paradigm of galaxy formation is built upon the foundation that all galaxies form within halos of dark matter (White \& Rees 1978; Blumenthal~et al. 1984). The evolution and growth of galaxies is governed by the merging of their dark matter halos and the conversion of gas into stars. The mass of a halo is therefore a fundamental parameter that influences a galaxy's form.

While the physics of halo growth in a cold dark matter (CDM) universe can be explored with computer simulations (e.g. Springel al.~2005), actually measuring the total dark mass content of galaxies is problematic.  Recently, great strides have been made in this area by utilising the technique of weak gravitational lensing to infer halo masses at very large radii (Tyson et al. 1984; Hoekstra~et al. 2005; Mandelbaum et al.~2006; Parker et al.~2007). This technique statistically combines the lensing signal from a large number of halos and can yield general relationships between galaxy stellar and total halo masses (Hoekstra~et al. 2005; Mandelbaum et al.~2006).  Such a relation also follows from analysis of the ``conditional luminosity function'', which is an observationally-based model connecting the stellar mass of galaxies to their hosting halos (Yang, Mo \& van den Bosch 2003, 2008; van den Bosch et al.~2007).  Individual galaxies are expected to scatter about these general relations. 

A number of other techniques exist for measuring the halo mass of {\it individual} galaxies: rotation curves in disk galaxies (Sofue \& Rubin 2001), X-ray gas in giant elliptical galaxies (O'Sullivan \& Ponman 2004), kinematics of globular clusters and planetary nebulae (Romanowsky et al. 2003; Romanowsky et al.~2008) and strong gravitational lensing (Ferreras, Prasenjit \& Williams 2005). However, such techniques all have their limitations, e.g. they are often only applicable to certain galaxy types, require equilibrium conditions, can be observationally very expensive, probe only a limited radial extent and, in the case of strong lensing, is only available for galaxies along random lines of sight.  As a result, only a very small number of individual galaxies have direct measures of their total halo mass available.

In this {\it Letter} we present a new method for estimating the total halo mass surrounding individual galaxies based on the total mass contained in their globular cluster (GC) systems.

All large galaxies, including the Milky Way, are known to host a system of GCs.  They are among the oldest ($>11$ Gyrs) stellar populations known (Strader et al. 2005; Puzia et al. 2005; Proctor et al. 2007) and therefore trace the first stages of galaxy formation (West et al. 2004; Brodie \& Strader 2006).  The total mass of a GC system should remain relatively constant over time because the dominant constituent of the GC system mass budget are the massive GCs, whose estimated lifespan is much longer than the age of the Universe (Mc~Laughlin 1999).  Furthermore, the overall properties of GC systems are essentially unaffected by galaxy evolutionary processes (e.g. quiescent, on-going star formation), which can sometimes dramatically change the global properties of a hosting galaxy.

Here we provide evidence for a direct correlation between the mass of a GC system and its host galaxy halo mass.  The direct proportionality implies that GC system masses can be used to approximate the halo masses of individual galaxies.

\section{Data}\label{data}

\begin{figure}
\resizebox{1\hsize}{!}{\includegraphics[angle=-90]{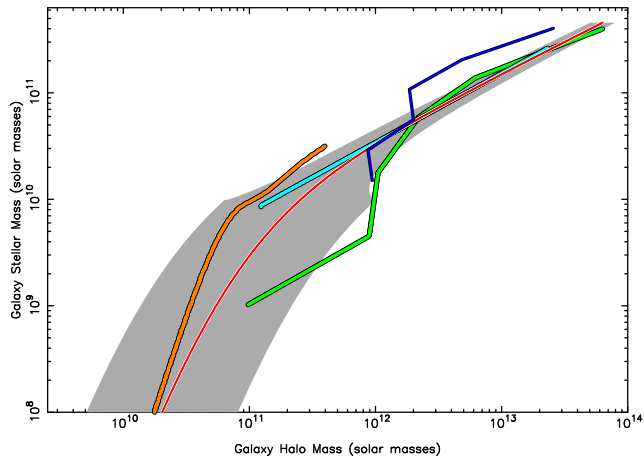}}
\caption{Statistical relations between galaxy halo and stellar mass. The green and cyan lines are from the weak gravitational lensing analysis of galaxies in isolated environments by Mandelbaum et al.~(2006) and Hoekstra et al.~(2005), respectively.  The blue line is the same for spiral galaxies (Mandelbaum et al.~2006).  The orange line comes from conditional luminosity function analysis (van den Bosch et al.~2007).  The adopted relation is the thin red line with 1-sigma errors shown as the grey region.}\label{fig1}
\end{figure}

\begin{figure}
\resizebox{1\hsize}{!}{\includegraphics[angle=-90]{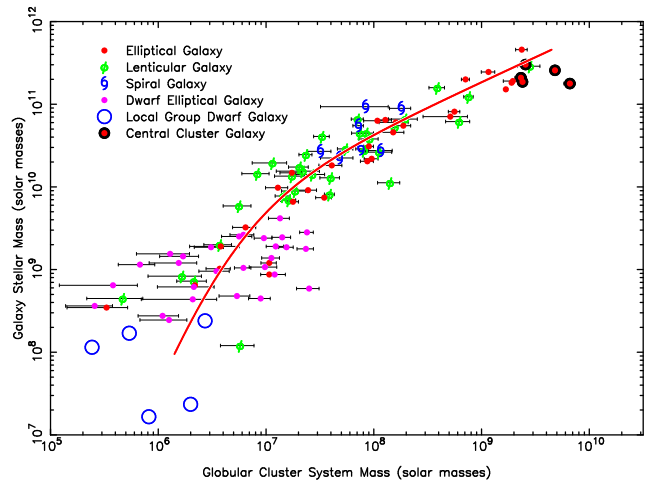}}
\caption{Galaxy stellar mass plotted against total globular cluster system mass.  The red line is the adopted relation in Fig.~\ref{fig1}, shifted in the X-axis by -4.15 dex from the direct proportionality between halo and stellar mass given in the text (Eq.~1).  It illustrates how galaxy halo and GC system masses are interchangeable.}\label{fig2}
\end{figure}

This section documents the data sources and how galaxy halo masses are approximated from their stellar masses.  Halo mass is defined as the total mass (baryon plus dark matter) within a sphere containing an over-density of 180 times the background (Mandelbaum et~al.~2006; van den Bosch et~al.~2007).  Literature virial masses are converted to the adopted definition of halo mass using Eq. 15~in van den Bosch et al.~(2007).  A Hubble parameter of $h = 0.7$ is adopted throughout.

As the Universe evolves, certain galaxies end up at the centre of very massive galaxy clusters, while others enter these cluster halos as satellite galaxies.  Halos associated with central galaxies can be significantly more massive than the halo around a satellite galaxy, even when the stellar masses are identical (Mandelbaum et al.~2006).  Thus a distinction between central and satellite galaxies must be made when applying statistical relations involving stellar mass.  Such a distinction has not been made in related work (Spitler et al.~2008; Peng et al.~2008).

The present sample of galaxies is dominated by satellites and galaxies that lie at the centre of halos much smaller than those hosting galaxy clusters.  To avoid the cluster-sized halos, which occupy regions of high galaxy number densities, we employ general relations between stellar and halo masses derived from {\it isolated environments}.  We assume that a halo in an isolated environment has a similar stellar mass to a satellite halo of the same halo mass.

Fig.~\ref{fig1} shows two empirical halo mass relations derived from weak galaxy-galaxy gravitational lensing analysis of galaxies in isolated environments (Mandelbaum et al.~2006; Hoekstra et al.~2005).  To supplement the relation for low mass galaxies, we use a restricted form of a relation derived by van den Bosch et al.~(2007) from analysis of the conditional luminosity function.  For this relation, only galaxies of $M_{stellar} < 5\times10^{10} M_{\odot}$ are relevant, because at higher masses, the relation of van den Bosch et al.~(2007) becomes dominated by the cluster-sized halos around central galaxies.

To use the Hoekstra et al.~(2005) and van den Bosch et al.~(2007) relations in our analysis, we had to convert their galaxy {\it luminosity} versus halo mass relation into one relating galaxy {\it stellar mass} to the halo mass.  Because our sample is dominated by massive elliptical and lenticular galaxies, we assume a mass-to-light ratio corresponding to a 12 Gyr stellar population and a metallicity of [Fe/H] $= +0.08$.  Although the low mass galaxies in our sample likely require smaller mass-to-light ratios because they tend to be younger and have lower metallicities, this effect is negligible ($\sim0.1-0.3$ dex) compared to the halo mass uncertainties for such galaxies.  

With the weak lensing and conditional luminosity function constraints, we can create a general relation between galaxy stellar ($M_{stellar}$) and halo masses ($M_{halo}$).  This is shown in Fig.~\ref{fig1} as the red line and is defined as:  \begin{equation} M_{stellar} = M_{s} { (M_{halo}/M_h)^{\alpha+\beta}\over{(1+M_{halo}/M_h)^{\beta}} } \end{equation}  This is the same form as used in Yang et al.~(2008), with scaling parameters: $log~M_s = 9.8$, $log~M_h = 10.7$, $\alpha = 0.6$ and $\beta = 2.9$.  Halo masses for the spiral galaxies in our sample were estimated using the spiral galaxy relation of Mandelbaum et al.~(2006), also shown in Fig.~\ref{fig1}.  Note the data in Fig.~\ref{fig2} show a similar form to the relationship adopted between galaxy stellar and halo mass (Fig.~\ref{fig1}), with only an offset in the X-axis values.

Galaxy stellar masses are taken from Spitler et al.~(2008) and Peng et al.~(2008).  For the Spitler et al.~(2008) estimates, the Chabrier (2003) initial stellar mass function is used to match the Mandelbaum et al.~(2006) relation.  There is a small (0.07 dex in $log~M_{stellar}$) systematic offset between the masses derived using the Spitler et al.~(2008) technique and those published in Peng et al.~(2008).  This offset is removed from the Peng et al.~(2008) masses before analysis.

The GC system numbers in Spitler et al.~(2008) were converted to GC system total masses by multiplying the numbers by the average GC mass of $4\times10^5$ M$_{\odot}$. Peng et al.~(2008) summed the total stellar mass of all GCs in each galaxy.  For galaxies in common, the Spitler et al.~(2008) GC masses were used because they come from wide-field imaging where the entire spatial coverage of the GC system was observed. The NGC 3311 GC system number estimate is from Wehner et al.~(2008).  For reference, a table is available online with relevant properties of the main sample.

The GC system mass of Local Group (LG) dwarf galaxies is estimated by summing the individual GC stellar masses inferred from V-band photometry (Harris 1996; Webbink 1985; Da~Costa \& Mould 1988) and applying a mass-to-light ratio of 2.2 for an old, metal-poor stellar population (Bruzual \& Charlot 2003).  LG dwarf galaxy stellar masses are from V-band absolute magnitudes (Lotz et al.~2004) with appropriate mass-to-light ratios from Bruzual \& Charlot (2003) for the age and metallicity of the stellar populations (Lotz et al.~2004).  Total masses ($M_{total}\propto\sigma_0^2$, where $\sigma_0$ is the central velocity dispersion) and distances to these galaxies are from Mateo~(1998).

The analysis includes five galaxy clusters selected because their central galaxy has a reliable GC system number measurement available.  GCs associated with galaxy clusters will reside in the central cluster galaxy, around satellite galaxies and in the intracluster medium (see \S\ref{results}).  The total mass of GCs associated with satellite galaxies was approximated by integrating the observed cluster galaxy mass functions (Sandage, Bingegeli, Tammann et al.~1985; Ferguson \& Sandage 1991; Yagi et al.~2002; Trentham, Tully \& Mahdavi 2006) after convolving them with a quadratic fit to data in Fig.~\ref{fig2}.  No global constraint on an intracluster GC population exists.  We therefore use a prediction from computer simulations of galaxy clusters (Bekki \& Yahagi 2006) that intracluster GCs make up 29\% (with RMS = 5\%) of the total cluster GC mass, independent of the clusters total mass.  Because the study of Bekki \& Yahagi (2006) was limited to a rudimentary GC formation prescription, formal uncertainties on these total cluster GC masses are taken to be 40\%.  Cluster halo masses are taken from the following sources: Virgo and Hydra clusters (Girardi et al. 1998), NGC~1407 (Brough et al. 2006), Antlia (Nakazawa et al. 2000), and Fornax (Drinkwater, Gregg \& Colless 2001).

\section{Results}\label{results}

\begin{figure}
\resizebox{1\hsize}{!}{\includegraphics[angle=-90]{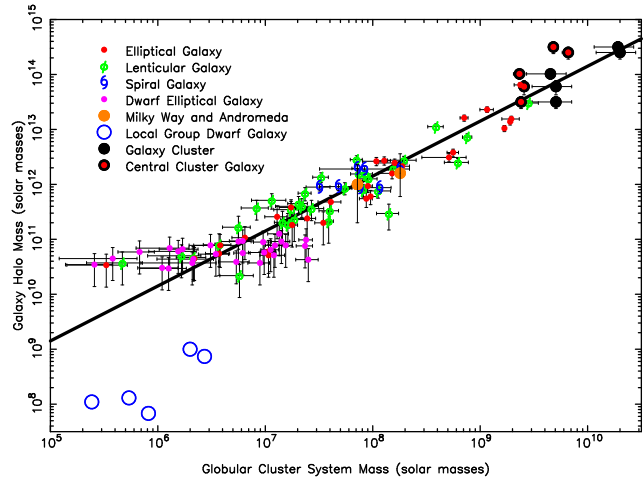}}
\caption{Galaxy halo mass plotted against total globular cluster system mass.  Galaxy halo masses are a measure of the total mass associated with a galaxy, including both dark and baryonic matter. The data points are consistent with the black line, which has a slope of unity.  This implies that GCs formed in direct proportion to the total halo mass of a galaxy: $\sim0.007\%$ (or $log~(M_{GCS}/M_{halo}) \sim -4.15$) of total halo masses are in the form of GCs.  Independent measures of the Milky Way and Andromeda halo masses are consistent with the relation.  Although the GCs immediately around central cluster galaxies apparently do not follow the relation (black-ringed red circles), the total GC mass associated with an {\it entire} galaxy cluster halo (including GCs around satellite galaxies and between them) does (black circles).  Local Group dwarf galaxies do not follow the trend at lower masses.  The apparent skewing of the main dataset at low masses is due to a sample bias in the galaxy stellar masses.}\label{fig3}
\end{figure}

\begin{figure}
\resizebox{1\hsize}{!}{\includegraphics[angle=-90]{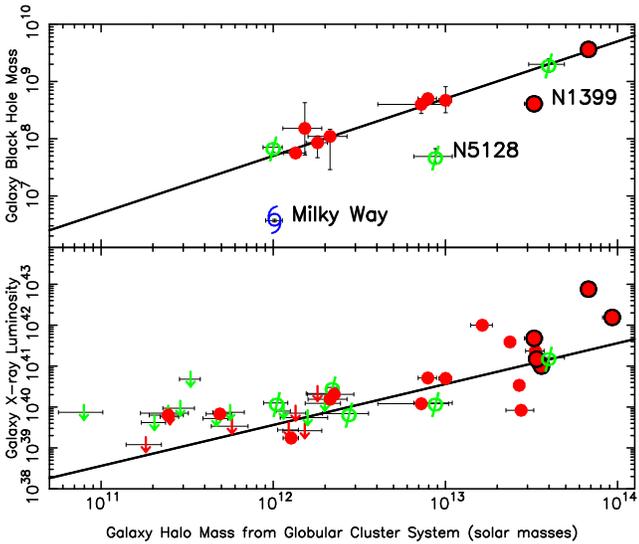}}
\caption{Two properties of galaxies plotted against the halo mass derived from their globular cluster systems and the relation of Fig.~\ref{fig3}. Symbols are as in Fig.~\ref{fig3}.  Arrows indicate upper limits. The black lines have a slope of unity.  {\it Top panel:} Galaxy supermassive black hole mass (units of $M_{\odot}$) from Graham~(2008a).  The data scatter about the unity line, suggesting that the central supermassive black hole in a galaxy is directly proportional to its halo mass.  Three outliers are evident, although this can be explained (see Gebhardt et al.~2007; Graham~2008b).  {\it Bottom panel:} Galaxy X-ray luminosities (units of $ergs~s^{-1}$) from O'Sullivan, Forbes \& Ponman~(2001).  The data are consistent with a direct proportionality between the X-ray luminosity of a galaxy and its halo mass.  Diffuse X-rays emitted by a galaxy are an indicator of the total mass of that galaxy.  Halo masses from GCs show a very tight relation with the X-ray luminosities.}\label{fig4}
\end{figure}

In Fig.~\ref{fig3} we show that GC system masses ($M_{GCS}$) are directly proportional to the total halo mass of its host galaxy, with a scatter comparable to the observational uncertainties.  The form of the line in Fig.~\ref{fig3} is $log~M_{halo} = log~M_{GCS}+4.15$.  This can be related to the initial total baryon mass of a galaxy, by assuming the universal baryon fraction (i.e. $M_{baryon}/M_{halo}\approx\Omega_{b}/\Omega_{m}\approx0.17$; Komatsu et al. 2008) applies on all galactic scales in the early Universe.

For galaxies with $M_{halo}>5\times10^{11} M_{\odot}$, the relationship in Fig.~\ref{fig3} appears to be invariant to the local environment and to the morphological type of the galaxy (see discussion in Spitler et al. 2008).  In contrast, the statistical relationships between stellar and halo mass depend on whether the galaxy is a spiral or an elliptical type (see Fig.~\ref{fig1}).  Furthermore, the statistical relationship between galaxy stellar mass and the halo mass is strongly non-linear.  This means that for very massive galaxies their halo masses derived from the stellar mass relation are poorly constrained.

The direct proportionality between GC system masses and their host halo is consistent with our current understanding of GC formation.  While stars in a galaxy can change with time and environment, the GCs remain relatively immune to the evolutionary processes operating on and within galaxies.  In the early Universe, GCs simply formed in proportion to the total amount of baryonic mass associated with a galaxy, which is directly related to its halo mass.

Computer simulations of GC system formation shows some disagreement in the relationship between GC numbers and halo masses (Kravtsov \& Gnedin 2005; Bekki et al. 2008).  It is noted that first self-consistent modelling of star cluster formation and evolution in a realistic galaxy potential was only recently accomplished (Hurley \& Bekki~2008).

Diemand, Madau \& Moore~(2005) and Moore et al.~(2006) assumed GC formation sites were associated with rare density peaks in the early cold dark matter density field and made some predictions relating the local epoch of reionisation to certain GC system properties.  In particular, if reionisation truncated GC formation throughout the Universe at roughly the same time, they predicted that the GC system numbers and hence masses should scale with the host galaxy halo mass.  In this context, our observations imply that, on average, the absolute difference between the local epoch of reionisation in galaxies of different masses and/or environment was shorter than typical GC formation timescales (see also discussion in Weinmann et al.~2007).

Two recent studies (Spitler et al. 2008; Peng et al. 2008) performed a similar analysis to that shown here, echoing past work with central cluster galaxies (Blakeslee et al. 1997; Mc~Laughlin 1999).  Although each study claimed evidence for a direct proportionality between GC number and the halo mass of the host galaxy, a comprehensive analysis has not been preformed until now.  Blakeslee et al. (1997) and Mc~Laughlin (1999) focused on very massive galaxies, but did not have access to total mass estimates for lower mass galaxies.  Spitler et al. (2008) preformed an initial analysis on a subset of the present sample and found the halo mass normalised GC numbers ($V_N$; Spitler et al. 2008) are constant.  This is in dramatic contrast to the trend of increasing relative GC system numbers when normalised by galaxy luminosities (i.e. the GC specific frequency, $S_N$).  Peng et al.~(2008) found similar results.  However, both Spitler et al.~(2008) and Peng et al.~(2008) employed a statistical stellar-halo mass relation contaminated with galaxy cluster-sized halos (see discussion in \S\ref{data}). 

We can test the robustness of the relation presented in Fig.1 by using independent halo mass measurements from direct observations.  Our own Milky Way Galaxy and the neighbouring spiral galaxy, Andromeda, have halo measurements of $1\times10^{12}$ and $2\times10^{12} M_{\odot}$, respectively (Xue et al.~2008; Lee et al.~2008).  As shown in Fig. 1, these independent halo masses are consistent with those expected from their GC system masses and therefore support the general relation between halo and GC system masses.  Confirmation of this relation will require larger samples of direct mass measurements of individual galaxy halos than are currently available.

Galaxies that reside at the centres of large clusters of galaxies are associated with an extremely massive halo. Five of these central galaxies are shown in Fig.~\ref{fig3} (as black-ringed, red circles), where the halo masses are direct measurements from observations (satellite kinematics and X-ray analysis, see \S\ref{data}). It is apparent that the central cluster galaxies deviate from the relation and show smaller GC system masses than might otherwise be expected for their halo masses.  However, these halos host numerous subhaloes (i.e. satellite galaxies) that also host GC systems.  Unlike the satellite galaxies of smaller halos (e.g. the Milky Way), the total GC mass of cluster subhaloes contributes a non-negligible fraction to the overall cluster GC mass.  Furthermore, galaxy clusters have a population of GCs not directly associated with individual galaxies, but are instead found between galaxies in the intracluster medium (West et al. 1995; Williams et al.~2007).  Thus when estimating the total GC system mass of a cluster halo, one should include GCs found in satellite galaxies and the intracluster medium, not just the central galaxy.

In Fig.~\ref{fig3}, the cluster-wide GC masses derived in \S\ref{data} are given as black circles.  These values are consistent with the relation for individual galaxies and further supports the idea that GC masses indeed trace the underlying halo mass irrespective of the physical scale. 

Five dwarf galaxies associated with the Local Group (LG) are included in Fig.~\ref{fig3}.  These galaxies have between 2--9 GCs and their total masses are derived from their central velocity dispersions ($\sigma_0$, see \S\ref{data}).  These galaxies do not follow the same trend established by massive galaxies.  This may reflect the gross extrapolation of the radial mass profile that is required to convert an observed $\sigma_0$ to a total mass.  If the LG galaxies actually follow the GC system--halo mass trend of the more massive galaxies, their GC system masses imply their halo masses range from $\sim10^{10}-10^{11}$.  This can be tested in the future with mass modelling of stellar kinematics at large radii (e.g. De Rijcke et al. 2006; Gilmore et al. 2007; Walker et al. 2007).  LG dwarfs without GCs (this is the majority) may fall below some halo mass threshold where massive GC formation is unlikely to have occurred.  

Recent observational efforts have also shown there exists an apparent difference in the halo masses of galaxies with similar stellar masses, depending whether or not its active galaxy nucleus (AGN) is ``radio-loud'' (Mandelbaum et al. 2008).  This observation can be tested with GC systems in future work.

\section{Applications}\label{apply}

The relation in Fig.~\ref{fig3} provides evidence that GCs formed in direct proportion to the total mass associated with a galaxy.  Hence by measuring the mass of a GC system we can infer the total halo mass of an individual host galaxy, enabling a comparison between the detailed properties of galaxies and their halo masses.

It is believed that most large galaxies harbour a supermassive black hole of $10^7$ to $10^{10} M_{\odot}$ in their nuclei.  Galaxies with well-studied GC systems and that have direct supermassive black hole mass measurements catalogued in Graham (2008a) are shown in Fig.~\ref{fig4}.  Except for 3 galaxies (see Fig.~\ref{fig4} caption) the sample shows a direct proportionality between black hole and halo mass, with very little scatter.  Supermassive black holes apparently show a close connection to the host galaxy halo mass (e.g. Ferrarese 2002; $c.f.$ Ho 2007).

In the lower panel of Fig.~\ref{fig4}, elliptical and lenticular galaxies with X-ray luminosity measurements from O'Sullivan, Forbes \& Ponman~(2001) are shown.  Gas of sufficiently high temperature and density will emit X-rays and the strength of the diffuse X-ray emission in a galaxy relates to its total mass. Although the X-ray measurements of the lower mass galaxies are mostly upper limits, the galaxy X-ray luminosities appear to directly correlate with the host halo mass inferred from their GC systems.  Interestingly, the scatter of the data in Fig.~\ref{fig4} (RMS $=0.47$ dex) is less than or equal to reported correlations between X-ray and optical luminosities (Ellis \& O'Sullivan 2006), suggesting GC system masses are indeed excellent tracers of galaxy halo masses.

Peng et al. (2008) reported a trend among dwarf galaxies where the average galaxy luminosity-normalised GC system numbers (i.e. the GC specific frequency, $S_N$) decreases with increasing radius from the central Virgo cluster galaxy, M87.  If the GC system numbers/masses reflect their host halo masses, the trend of Peng et al. (2008) can be interpreted as an increasing field star formation efficiency with increasing cluster-centric radius.

The notion that GCs trace the total mass of galaxies may apply on sub-galactic scales.  Blakeslee et al. (1997) and Mc~Laughlin (1999) provide evidence for this in the central regions of very massive elliptical galaxies.  If this holds for lower-mass galaxies as well, GC system surface density profile could trace the underlying dark matter halo mass profile.

Finally, GC systems generally show two colour subpopulations, which are interpreted as evidence for two metallicity subpopulations of metal-poor and metal-rich GCs (see discussion in Spitler, Forbes \& Beasley 2008).  Metal-poor GCs are thought to have formed during a pre-galactic era, while the metal-rich GCs more closely relate to a galaxy's field star population (Brodie \& Strader 2006).  It follows that metal-poor GCs might show a closer relationship to the host halo mass.  In our sample, metal-poor GCs in all galaxies more massive than $M_{halo}>5\times10^{11} M_{\odot}$ ($M_{stellar}>10^{10} M_{\odot}$) make up roughly $\sim60\%$ (RMS $=5\%$) of its GC system.  This percentage increases in lower mass galaxies so by $M_{stellar}\sim10^{9} M_{\odot}$ most GC systems have $\sim90\%$ metal-poor GCs.  However, the halo masses of these lower mass galaxies are poorly constrained, thus we currently cannot determine whether the metal-poor GCs show a closer relationship to its host halo.

\section{Conclusions}

We have presented evidence for a direct correlation between the total GC system mass and the host galaxy halo mass.  While this result appears to be robust for intermediate-to-massive galaxies (and possibly galaxy clusters), the halo masses of low-mass galaxies are not well-constrained, and thus the correlation at low galaxy masses must be confirmed with better halo mass estimates.

The direct correlation implies GC system masses can be used to directly measure the total halo mass of individual galaxies.  This technique has the advantage that it can be applied irrespective of the host galaxy type and environment.  It is relatively inexpensive in terms of observing time, requiring only wide-field imaging under reasonable observing conditions (e.g. Spitler et al.~2008).  A few examples have been demonstrated here to help illustrate the promising astrophysical applications of this new technique.

\section*{Acknowledgments}

We thank K. Bekki, K. Glazebrook, A. Graham, R. Mandelbaum, J. Strader, and A. Romanowsky for reviewing a draft of the manuscript.  We are grateful for useful discussions with R. Mandelbaum and T. Mendel.  We also thank A. Graham for use of his supermassive black hole catalogue.  DAF acknowledges the ARC for financial support.  The referee, B. Moore, provided a number of comments that enhanced the discussion.

\label{lastpage}

\end{document}